Type of the Paper (Review)

# New horizon in the statistical physics of earthquakes: Dragon-king theory and dragon-king earthquakes

Jiawei Li [1], Didier Sornette [1*], Zhongliang Wu [1,2], Hangwei Li [1]

1 Institute of Risk Analysis, Prediction and Management (Risks-X), Academy for Advanced Interdisciplinary Studies, Southern University of Science and Technology (SUSTech), Shenzhen 518055, China
2 Institute of Earthquake Forecasting, China Earthquake Administration, Beijing 100036, China

**Abstract**：A systematic quantitative investigation into whether the mechanisms of large earthquakes are unique could significantly deepen our understanding of fault rupture and seismicity patterns. This research holds the potential to advance our ability to predict large earthquakes and enhance the effectiveness of disaster prevention and mitigation strategies. In 2009, one of us introduced the dragon-king theory, offering a quantitative framework for identifying and testing extreme outliers—referred to as dragon-king events—that are endogenously generated. This theory provides valuable tools for explaining, predicting, and managing the risks associated with these rare but highly impactful events. The present paper discusses the feasibility of applying this theory to seismology, proposing that dragon-king earthquake events can be identified as outliers to the Gutenberg-Richter law. It also examines several seismological mechanisms that may contribute to the occurrence of these extraordinary events. Although applying the dragon-king theory to seismology presents practical challenges, it offers the potential to significantly enrich statistical seismology. By reexamining the classification of earthquake rupture types through a statistical testing lens and integrating these insights with underlying physical mechanisms, this approach can greatly enhance the analytical tools and depth of research in the field of statistical seismology.

**Keywords**：statistical physics of earthquakes; dragon-king event; dragon-king theory; earthquake mechanism; statistical seismology; outlier detection

## 0 Introduction

In an average year, the Earth typically experiences one great earthquake with a magnitude 8 or higher, along with approximately 10 to 15 major earthquakes measuring 7 or above, around 150 strong earthquakes with magnitudes 6 or higher and so on. While large earthquakes occur infrequently, the immense energy they release within mere seconds to minutes wields significant influence over the Earth's crustal energy budget. This rapid energy release contrasts sharply with the prolonged tectonic processes involved in stress accumulation and with geological timescales. Large earthquakes can cause widespread effects, as observed with the 1976 Hebei Tangshan $M_W$ 7.6 / $M_S$ 7.8 earthquake (Chen et al., 1988; Mearns and Sornette, 2021), the 2008 Sichuan Wenchuan $M_S$ 8.0 earthquake (Chen and Booth, 2011), and the 2011 Japan Tohoku $M_W$ 9.0 earthquake (Satake et al., 2013).

The outstanding question we discuss in this review is whether large earthquakes enjoy some distinguishable properties that could make them predictable. This question is motivated both by the quest of understanding the physics of earthquakes and their spatio-temporal organization and by the profound implications this would have for seismic risk assessment and mitigation.

**(1) Viewpoint: No!**

This research was supported by the Guangdong Basic and Applied Basic Research Foundation (Grant No. 2024A1515011568), the National Natural Science Foundation of China (Grant no. U2039202 and U2039207), Shenzhen Science and Technology Innovation Commission (Grant no. GJHZ20210705141805017), and the Center for Computational Science and Engineering at Southern University of Science and Technology.

*Corresponding author. Sornette D, email: didier@sustech.edu.cn

There is a widespread view that the answer to this question is negative: "large earthquakes are just like small earthquakes that continued to grow to large sizes." In this picture, large earthquakes are not different from their smaller siblings, except for their sizes, with scaling relations describing the statistical properties of the full population of earthquakes. This perspective is informed by the many power law distributions characterizing seismicity statistics.

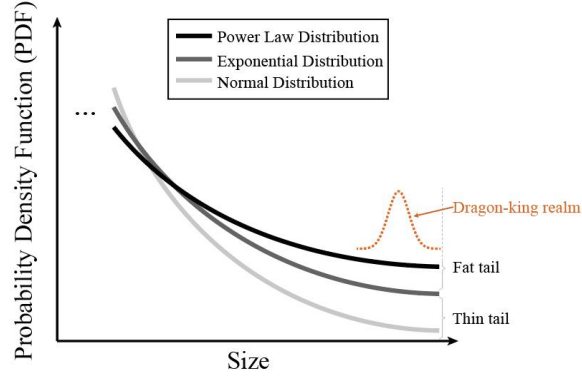

Fig. 1 A schematic diagram of the observations of dragon-king events realm appearing as outliers in the tail of a power law distribution. Typically, if the probability density function decays faster than an exponential distribution in the tail, it is referred to as a thin tail, such as for a Gaussian distribution; whereas a decay slower than an exponential is known as a fat tail. A specific example of a fat tail distribution is the power law distribution. The term "heavy tail" is technically used to refer to an asymptotic power law decay with exponent $\alpha$ defined in Equation (1) smaller than 1, so that the variance of the distribution is mathematically infinite. The dragon-king regime can express itself as a shoulder or as a change of speed of decay in the far-right end of the distribution.

Power law distributions are quite ubiquitous in complex systems across natural sciences, sociology, and economics (Mitzenmacher, 2004; Newman, 2005; Sornette, 2006), describing the probability density function $P(x)$ of event size $x$ as

$$P(x) \propto \frac{1}{x^\alpha} \tag{1}$$

where $\alpha$ is a positive constant exponent. Power law distributions exhibit two key properties: self-similarity and fat tails. Self-similarity implies that all events in the system, including extreme large and small events, are identical in their generating mechanisms. Fat tails indicate that the probability of extreme large events occurring at the tail of the distribution far exceeds that of exponential and normal distributions (Figure 1; Mitzenmacher, 2004; Newman, 2005; Sornette, 2006). Power law distributions that approximately describe many observations in seismology include the Gutenberg-Richter law of the frequency of earthquake energies or seismic moments (when expressed with magnitudes, the Gutenberg-Richter law is an exponential distribution) (Gutenberg and Richter, 1944; Sornette et al., 1996; Liu et al., 2024), the frequency distribution of fault rupture scales (Davy, 1993; Scholz et al., 1993; Scholz, 1997; Clark et al., 1999), the fractal/multifractal/hierarchical scaling features of seismicity and fault systems (Hirata and Imoto, 1991; Turcotte, 1997; Hirabayashi et al., 1993; Ouillon et al., 1996), the frequency distribution of seismicity rates in spatiotemporal grids (Saichev and Sornette, 2006; 2007), the frequency distribution of the time intervals between large earthquakes and their epicentral distances (Ito, 1995), and the frequency distribution of earthquake fault slip (Lavallée and Archuleta, 2003; Lavallée et al., 2006), among others. The evidence for self-similarity between earthquakes is further reinforced by the existence of several approximate seismicity laws also formulated as power laws. To a first-order approximation, the Omori-Utsu law describes the decay rate of aftershock activity following a mainshock (Omori, 1894; Utsu, 1970), the fertility law quantifies the exponential dependence of the average number of aftershocks triggered by a given mainshock as a function of its magnitude (it is thus a power law function of the earthquake energy) (Yamanaka and Shimazaki, 1990), the spatial Green function specifies the decay of the number of aftershocks with distance from the mainshock's epicenter (Zhuang et al., 2004; Nandan et al., 2021), and the $k^{-2}$ spectrum in the wavenumber domain describes the relative content of spatial frequency components of earthquake fault slips (Herrero and Bernard, 1994), among others.

The widespread presence of power law distributions in seismology (especially the Gutenberg-Richter law) has been interpreted as supporting the concept of self-organized criticality in seismology (Bak et al., 1987; Bak and Tang, 1989; Sornette and Sornette, 1989; Bak and Paczuski, 1995), which holds that large earthquakes are not essentially special compared to other earthquakes; all events reveal a self-organization of the Earth crust to a critical state characterized by slow stress built-up released by sudden earthquakes of all sizes (Geller et al., 1997; Wyss, 1997). The fact that both large and small earthquakes follow the same approximate power law distribution is interpreted as implying that they have the same nucleation mechanisms, same seismic genesis, and same rupture mechanisms. If true, this would make it impossible to identify truly informative precursors before a large earthquake, thereby rendering large earthquakes unpredictable. This perspective has been articulated in details in the discussion forum organised in the journal Nature in 1999 on whether earthquakes can be predicted (https://www.nature.com/nature/articles?searchType=journalSearch&sort=PubDate&type=debate&year=1999&page=4. Last accessed May 10, 2024).

**(2) Viewpoint: Yes!**

In contrast to this view of perfect self-similarity, other observers have noted that, in many specific regions (such as plate boundary areas or Late Quaternary active fault zones), the size-frequency relationship of earthquakes significantly deviates from the Gutenberg-Richter law at the large earthquake end. This has led to the characteristic earthquake model (Wesnousky, 1994; Ishibe and Shimazaki, 2012) according to which large faults can rupture episodically with earthquake ruptures spanning their whole length. There is also some evidence that large and small earthquake groups follow two different distributions in 14 subduction zones of the Circum-Pacific Seismic Belt (Pisarenko and Sornette, 2004), where large here means $M_W > 8.1 \pm 0.3$, a magnitude range usually referred to as corresponding to "great" earthquakes. This suggests that earthquakes with $M_W > 8.1 \pm 0.3$ may have special properties compared to smaller ones. Additionally, computer simulations using boundary integral methods to solve the rupture dynamics on homogenous faults based on the physics of elasto-dynamical frictional sliding show that there may be significant differences in the nucleation and rupture propagation of large and small earthquakes, beyond just their size differences (Wen et al., 2018; Wei et al., 2021; Xu et al., 2023; Sornette et al., 2024). Specifically, there is a class of earthquakes that self-arrest shortly after the initial rupture begins, without any external barriers (Wei et al., 2021). It is estimated that these self-arresting earthquakes have a maximum magnitude of about 6 to 6.5. In addition to their discovery in careful numerical simulations, such self-arresting earthquakes have also been observed in earthquake catalogues (Galis et al., 2017; Wen et al., 2018; Xu et al., 2023). Computer simulations also show that the self-arresting earthquakes coexist with a class of run-away unstable ruptures (i.e., sub-shear and super-shear ruptures), but in different stress and frictional regimes. These run-away unstable ruptures are more likely to be large earthquakes because strong barriers are needed to stop their rupture propagation. Sornette et al. (2024) recently proposed a statistical thermodynamics approach for understanding the nucleation of earthquakes on homogeneous faults, clarifying the mechanisms behind the occurrence of these different types of earthquakes. By linking the fractal characteristics of fault networks with nucleation physics, Sornette et al. (2024) derived that self-arresting earthquakes satisfy a Gutenberg-Richter power law distribution of size. In this theory, the Gutenberg-Richter law is interpreted as being a statistical property of self-arresting earthquakes, while run-away events should have different distributions. These findings suggest that the genesis of large earthquakes may involve amplification mechanisms distinctly different from those of smaller earthquakes, possibly accompanied by observable precursors. A systematic and quantitative examination of whether there are essential differences (especially in terms of seismic genesis mechanisms) between small and large earthquakes not only could aid in better understanding the patterns of fault rupture and seismicity but also could have significant implications for research on earthquake predictability, disaster prevention and mitigation practices in scenarios involving large earthquakes (e.g., Li et al., 2022; Zhang et al., 2023).

It must be recognized that current seismology still lacks a usable analytical framework to systematically and quantitatively examine whether large earthquakes are special. Fifteen years ago, one of us introduced the concept of dragon-king events and formulated the dragon-king theory, which provides a potential, effective statistical quantitative analysis tool for discussing the above issues. The present paper will first introduce the definition of dragon-king events as applied to the description of extreme events occurring in complex systems and articulate the main elements of the dragon-king theory. We will then discuss the possible physical mechanisms for the emergence of dragon-kings in seismicity. We will present the main methods to identify dragon-king earthquakes in seismology, and finally point out several issues that need attention and continuous exploration for a practical application of the dragon-king theory to seismology.

# 1 Dragon-king events and dragon-king theory

## 1.1 Definition of dragon-king events

Dragon-kings (DK) embody a dual metaphor, expressing that an event is both exceptionally large ("king") and originates from unique circumstances ("dragon") compared to its peers. The hypothesis proposed by Sornette (2009), with a premonitory discussion emphasizing the predictability of catastrophes that are outliers (Sornette, 1999), posits that DK events arise from distinct mechanisms that occasionally amplify extreme events, resulting in the creation of runaway disasters on the downside as well as extraordinary opportunities on the upside. Possible generative mechanisms, statistical characteristics, modeling predictions, and risk management strategies of dragon-king events have been explored in (Sornette, 2009; Sornette and Ouillon, 2012; de S. Cavalcante et al., 2013).

The motivation for introducing the concept of dragon-kings came first from the empirical observations of profits and losses in stock markets, specifically that the frequency of extreme financial drawdowns (cumulative losses from peak to valley) is 10 to 100 times larger than predicted by the extrapolation to extreme values of the frequency distribution describing the vast majority of drawdowns (Johansen and Sornette, 1998; 2001). In other words, extreme financial drawdowns are "outliers": they lie much above the frequency distribution calibrated on 99% of the drawdown population. But, in statistics, outliers refer usually to unwanted data points. They are errors or data anomalies that need to be corrected or removed (National Institute of Standards and Technology, 2007). In contrast, the extreme outliers identified in the frequency distribution of financial drawdowns are not errors. In fact, they represent the most important events of financial markets, namely the major crashes that destroy trillions of USD in value in the short span of a few days or weeks. These extraordinary events impact the general economy and often trigger drastic responses from policy makers as well as the emergence of new regulations. Moreover, these extreme events have been found to be often the result of endogenous maturation processes towards instabilities. Thus, to distinguish them from outliers that are often of exogenous origin, the concept of dragon-kings was formulated to highlight the overwhelming significance of these events, emphasize their endogenous origin and draw attention to their unique and substantial impact (Sornette, 2009).

To date, dragon-king events have been widely reported in many studies, including:

1) Natural sciences, such as material failure, earthquake ruptures (Sammis and Sornette, 2002), fluid dynamics (L'vov et al., 2001), landslide prediction (Lei et al., 2023), epileptic seizures (Osorio et al., 2010; de Arcangelis, 2012);

2) Finance, such as stock market crashes (Johansen and Sornette, 1998; 2001; 2010; Filimonov and Sornette, 2015);

3) Sociology, such as the impact of academic papers (Golosovsky and Solomon, 2012), city sizes (Wheatley and Sornette, 2015), and nuclear safety accidents (Wheatley et al., 2017), among others.

## 1.2 Potential mechanisms of dragon-king events

Dragon-king events arise from unique amplification mechanisms that do not affect (or affect to lesser degree) other events in the same system, thus endowing dragon-king events with their inherent specific properties. To date, several mechanisms have been proposed to explain the occurrence of dragon-king events:

**(1) Exogenous shocks**. Most natural and social systems are continuously subjected to external stimulations, noises, shocks, and different types of forcing, which widely vary in amplitude. A very large event can thus be due to a strong exogenous shock, as opposed to the internal dynamics of the system, or maybe to a combination of both. A striking example is the massive impact on the Earth biosphere caused by the meteorite strike 66 million years ago, which is associated with the extinction of the dinosaurs. It is exogenous to the Earth biosphere and an extreme event of extraordinary impact.

An interesting question is to explore how does the presence of these exogenous dragon-kings alter the various statistical metrics that characterise the system's dynamics. In the context of the general hierarchical Weierstrass-Mandelbrot continuous-time random walk (WM-CTRW) model, Werner et al. (2012) answered to this question by inserting manually either a dragon-king in the duration of trends or a shock dragon-king whose size dwarfs any existing structure and they determined analytically and numerically the corresponding changes of the stationary velocity autocorrelation function.

**(2) Partial global synchronization.** Many systems can be schematically represented as coupled heterogeneous oscillators of relaxation with thresholds. Examples include biological neuron network exhibiting avalanches of neuronal firing, fault networks supporting earthquakes, cardiac cells, ecosystem populations, social systems, power grids, climate systems and so on. Different regimes occur depending on the ratio $C/H$ of the coupling strength $C$ to the magnitude $H$ of heterogeneity (for instance in the natural frequencies of the oscillators). For large $C/H$, the systems tend to be globally synchronized as illustrated by certain species of fireflies in Southeast Asia where large groups of male fireflies flash their bioluminescent light in unison to attrack females. For small $C/H$, the oscillators are oscillating approximately independently and the systems exhibit noisy spatio-temporal dynamics. At intermediate values of $C/H$, the systems present different forms of self-organisation with avalanches distributed over broad ranges of sizes. When the distribution of avalanche sizes is well-described by a power law, such systems are then referred to as self-organized critical (SOC). In between the SOC and fully global synchronized regimes, systems exhibit a mixture of power law-like distributions of avalanches together with transient bursts of global synchronization. These bursts are dragon-kings compared with the rest of the avalanche population described by power law distributions. These dragon-kings result from failed attempts of the system to fully enter the globally synchronized regime. As an example, the system studied by Gil and Sornette (1996) of coupled subcritical bifurcating systems, interconnected through the diffusion of a distributed control parameter, is represented by ordinary differential equations coupled with a stochastic partial parabolic diffusion equation. This system demonstrates both the self-organized criticality and dragon-king regimes depending on the time scale of the relaxation of avalanches. Similar behaviors have been studied in spring-block systems (Sachs et al., 2012), forests that generate wildfires (Sachs et al., 2012), neuron networks (de Arcangelis et al., 2012) and biological neuron networks in which large epileptic seizures can be considered as dragon-kings (Osorio et al., 2010).

**(3) Clustering behavior and percolation phase transitions.** The percolation problem studies the behavior and properties of connected clusters in arbitrary graphs. Percolation theory offers insights into a wide range of phenomena, providing a principled methodology to study, for instance, the properties of porous materials in Material Sciences, to model the spread of diseases in Epidemiology and to characterise the robustness of networks in Network Theory. As the fraction $p$ of connected sites or of existing links increases, a well-defined percolation critical transition occurs at a topology-dependent critical value $p_c$ above which a large-scale connectivity emerges. For $p > p_c$, a giant connected cluster appears that allows for percolation through the system. The giant component is a dragon-king among all clusters, which are otherwise distributed in size according to a truncated power law distribution (Stauffer and Aharony, 2018). There are many generalization of the percolation problem in which the presence of links between elements exhibit spatial correlation or inter-dependences of various types. In particular, rupture in disordered media can be conceptualized as belonging to the general class of "correlated percolation" problems, with the giant cluster being the crack network that is responsible for the system failure. In all these systems, the DK correspond to the giant component or rupture network or failure surface. See Chapter 13 in (Sornette, 2006). Controlling percolation via various schemes also often leads to the abrupt appearance of very large clusters (Schröder et al., 2017).

In the theory of Bose-Einstein condensation transitions, Bose-Einstein condensate droplets exhibit a supercritical clustering mechanism due to the relative attraction or utility control of their largest entities, leading to clusters larger than other droplet sizes (i.e., dragon-king events). Similar to the Bose-Einstein statistical problem of distributing $N$ particles among $C$ energy levels, Yukalov and Sornette (2012) considered the distribution of $N$ individuals among $C$ cities when studying the attractiveness of city sizes. They found that, just like Bose-Einstein condensation, populations also tend to cluster in the largest city under certain conditions, causing the size of the largest city to deviate from the Zipf distribution of other city sizes. Since Bose-Einstein condensation is a phase transition, the emergence of dragon-king events can also be considered a type of phase transition. Additionally, in studies of grand canonical minority games—where $N$ individuals compete for limited resources by choosing among $C$ strategies ($C < N$)—Johnson and Tivnan (2012) found that, when a strategy proves successful, a large number of individuals are attracted to adopt this strategy, leading to an extreme change in the system, which is associated with deviations from the standard statistics of the rest of the dynamics, i.e., dragon-kings. However, this change is transient and unsustainable, as more and more individuals switch to this previously successful strategy, which quickly becomes ineffective in the game.

**(4) Positive feedback mechanism**. Positive feedback refers to a process where an initial change is amplified by the subsequent reactions it provokes, leading to a further increase in the effect (Sornette, 1999; Sammis and Sornette, 2002). This can create a self-reinforcing loop where the initial stimulus is continuously intensified. In other words, an initial deviation from some starting point is followed by increasing deviations, in contrast with a negative feedback process that would bring back the system to the initial point. In economics and finance, positive feedback is called procyclicality, which is characterized by actions that strengthen economic trends, whether positive or negative, thus increasing the volatility of economic cycles. There are many examples of positive feedback processes in action in different domains. In climate dynamics, as the Arctic ice melts, it reduces the albedo effect (the reflection of sunlight), leading to more heat absorption by the Earth's surface, which further accelerates ice melt. In social systems, the value of a social media platform increases as more people join and use it, which attracts even more users, enhancing its network value. In financial markets, during a rising market, increasing stock prices attract more investors, driving prices higher, which attracts even more investors. Conversely, during a losing market, falling prices cause more selling, which further depresses prices potentially leading to a "death spiral", a financial crash. The increase in the number of running buffaloes intensifies the panic within the herd, which in turn prompts more buffaloes to join the run. Similarly, a bank run occurs when a large number of a bank's customers withdraw their deposits simultaneously due to fears that the bank might become insolvent. As more customers rush to the bank to withdraw their money, the fear of other customers concerning the future solvency of the bank increases, which amplifies the initial withdrawal behavior, leading to a self-reinforcing cycle of increasing withdrawals and a self-fulfilling prophecy of the bank run.

Unbridled, positive feedbacks generally leads to finite-time singularities, i.e., the explosion to very large values or the collapse to zero in finite time. For example, consider a bar with cross-section $A$ subjected to a constant tensile force $F$. A creep/damage process occurs within the material of the bar, which is captured by the phenomenological equation $dA/dt = -(F/A)^n$, where $F/A$ is the stress within the bar and $n$ is a positive exponent. The solution of this equation is $A(t) = A_0 (t_c - t)^{1/(n+1)}$, showing that the bar ruptures in finite time at $t = t_c$ and the stress correspondingly diverges in finite time. During this creep and damage process, many small cracks nucleate and merge, leading to detectable acoustic emissions. The final rupture in general exhibits a burst of energy release, associated with the underlying mathematical finite-time singularity (Gluzman and Sornette, 2001; 2002) that is many times larger than all previous acoustic emissions, qualifying a dragon-king event (Johansen and Sornette, 1999; Sornette, 2009).

Positive feedback mechanisms lead to dragon-kings, obtained as the large final events of a non-sustainable dynamics. The mathematical finite-time singularity is never fully realized because, close to the singularity, other forces and mechanisms come into play, the most obvious one being the finite size of the system. The divergence is replaced by extreme fluctuations or events, which are the dragon-kings.

A first illustration is provided by the distribution of citations of scientific papers. In the presence of linear preferential attachment, the distribution of citation numbers over the ensemble of scientific papers is found to be a power law (1). In the presence of superlinear preferential attachment, there is a runaway behavior where a few papers, the dragon-kings, contain a finite fraction of all citations (Krapivsky et al., 2000; Dorogovtsev et al, 2000), akin to a Bose-Einstein condensation mentioned above. Similarly, Golosovsky and Solomon (2012) found that highly cited papers (cited more than 1500 times) exhibited exceptionally large deterministic growth beyond random parts, also displaying dragon-king event characteristics. Even a weak memory of the past dynamics and its feedback on the rate of the multiplicative random walk may lead to divergent statistical distribution and runaways.

Table 1. A few statistics that can be used to detect dragon-king events. Consider a set of $n$ random numbers that have been ordered in descending rank $x_1 > x_2 > ... > x_n$. The different statistics are obtained as different combinations of sums and denominators.

| Statistics | Definition | Masking | Swamping | Tips |
|---|---|---|---|---|
| Standard block test SS statistic[1] | $T_k^{\text{SS}} = \dfrac{\sum_{i=1}^{k} x_i}{\sum_{i=1}^{n} x_i}$ | Numerator none | Present | Can detect multiple outliers |
| Improved block test SRS statistic[2] | $T_{k,r}^{\text{SRS}} = \dfrac{\sum_{i=1}^{k} x_i}{\sum_{i=r+1}^{n} x_i}$ | None | Present but decreased significantly | -- |
| Standard inward test MS statistic[3] | $T_j^{\text{MS}} = \dfrac{x_j}{\sum_{i=j}^{n} x_i}$ | Present | None | Not as good as SS and SRS when handling multiple outliers |
| Improved inward test MRS statistic[4] | $T_{j,r}^{\text{MRS}} = \dfrac{x_j}{\sum_{i=r+1}^{n} x_i}$ | Present but decreased significantly | None | -- |
| D statistic[5] | $T_k^{\text{D}} = \dfrac{x_1}{x_{k+1}}$ | Severe | Minor | -- |
| Block test DK statistic[6] | $T_k^{\text{DK}} = \dfrac{\sum_{i=1}^{k} z_i}{\sum_{i=k+1}^{n} z_i}$ | Severe | Severe | Unable to detect multiple outliers |

[1] SS statistic: sum-sum test statistic;
[2] SRS statistic: sum-robust-sum test statistic, in which $r$ is the pre-specified the maximum number of candidate outliers ($r \geq 1$);
[3] MS statistic: max-sum test statistic for the $j$th rank;
[4] MRS statistic: max-robust-sum test statistic for the $j$th rank;
[5] D statistic: Dixon test statistic;
[6] DK statistic: dragon-king test statistic (Pisarenko and Sornette, 2012), where for $i = 1, 2, ..., n - 1$, $z_i = i(x_i - x_{i+1})$ and $z_n = nx_n$.

A second illustration is found in the distribution of financial drawdowns, the peak to valley losses (with some tolerance of volatility during the descent), which exhibits very neat outliers. These extreme drawdowns are the large well-known financial crashes (Johansen and Sornette, 2001; 2010; Filimonov and Sornette, 2015). The positive feedbacks occur in two stages: (i) during the pre-crash bubble, prices develop along a super-exponential acceleration fueled by imitation, herding as well as procyclical hedging and many other potential positive feedback mechanisms (Sornette and Woodard, 2009; Sornette and Cauwels, 2015); (ii) during the crash, positive feedback loops create a downward spiral, where falling prices trigger more sales and the increase selling pressure drives prices even lower.

*1.3 Methods for detecting dragon-king events*

Since the statistical signature of dragon-king events is primarily that they are outliers, the current technical approach to identifying dragon-king events mainly involves testing outliers relative to a null model, i.e., a model without outliers, which in many cases assumes an underlying Gaussian distribution. However, empirical data in many fields often do not follow Gaussian distributions. In many scenarios, data can be better described by distributions with fatter tails such as exponential or power law distributions. In particular, Pareto (power law) distributions are prevalent across a broad spectrum of phenomena, ranging from natural hazards like earthquakes, landslides, floods, and tsunamis, to industrial catastrophes such as chemical spills, nuclear accidents, and power blackouts, and extend to social systems and geopolitical events, including the distribution of wars and conflict intensities measured by human losses (Laherrere and Sornette, 1998; Mitzenmacher, 2004; Newman, 2005; Sornette, 2006).

Following Wheatley and Sornette (2015), we focus on the detection of outliers in samples having approximately exponential or Pareto tails. Notably, through a simple logarithmic transformation, outlier tests designed for exponential samples can also be applied to Pareto samples. This places the test statistics with exponential underlying distribution at the center. There are four broad categories of tests: (i) block tests, (ii) inward tests, (iii) outward tests and (iv) mixture tests. Each of these tests are differently impacted by masking and swamping. Masking refers to the scenario where the next larger outlier can only be correctly detected if the previous outlier does not exist. In other words, successfully identifying a previous outlier may lead to missing the detection of the next outlier. Swamping applies to scenarios where a non-outlier is mistakenly detected as an outlier when considered alongside true outliers. In this case, the presence of true outliers can lead to false alarms among data points that are not actual outliers.

More specifically, let us assume there are $n$ ranked random variables $x_1 > x_2 > ... > x_n$, among which are found $k$ true outliers, $x_1, x_2, ..., x_k$, and $r$ candidate outliers to be tested with $n \geq r \geq k$). The null hypothesis $H_0$ assumes there are no outliers, meaning all variables originate from the same distribution. The four categories of outlier tests can be described as follows.

**(1) Block Test**. The block test is a direct method for identifying outliers, which determines all $r$ or 0 outliers in a single test by pre-specifying the number of candidate outliers, $r$. This method can suffer from masking and swamping if too many or too few data points are included within block of size $r$. However, when properly set, considering that the block test can handle all data simultaneously, this method still has strong practicality.

**(2) Inward Test**. The inward and outward sequential testing methods were developed to reduce reliance on the block size $r$. The inward test starts with the full sample $n$, and tests whether the largest data point $x_1$ is an outlier. If $x_1$ is identified as an outlier (i.e., the null hypothesis is rejected), it is removed from the sample of $n$ points, and the test is repeated for $x_2$. This process continues until the null hypothesis is not rejected for the first time. Thus, the estimated number of outliers is the number of rejected (marginal) tests. This test can be affected by both masking and swamping.

**(3) Outward Test**. This method starts by removing $r$-1 candidate outliers from the sample $n$ points, retaining only the smallest of the $r$ candidate outliers to be tested, $x_r$, to examine if it is a true outlier and all other $r$ - 1 larger candidate outliers are then qualified also as outliers. If the null hypothesis is not rejected, the algorithm performs on step outward to include the next smallest of the $r$ candidate outliers, $x_{r-1}$, back into the sample for testing. The process stops once the null hypothesis is rejected for the first time. If after traversing all $r$ candidates the null hypothesis is never rejected, then no outliers are identified. This testing method minimizes the likelihood and degree of masking and swamping.

**(4) Mixture Test**. This testing method is based on fitting the data to the following mixure model,

$$f(x) = (1 - \omega)\phi(x) + \omega\psi(x) \tag{2}$$

where $\omega$ is a weight factor ($0 \leq \omega \leq 1$), $\phi(x)$ is the distribution that all samples under the null hypothesis satisfy, and a Gaussian density function $\psi(x)$ with mean $\mu$ and standard deviation $\sigma$ defines the dragon-king realm. The null hypothesis $H_0$ is that $\omega = 0$. Using this model and the null hypothesis $H_0$ (a pure exponential or power law distribution) for a likelihood ratio test provides a *p*-value, while $n\omega$ estimates the number of outliers identified. This method does not require sequential testing. This method can be generalized to other forms for the dragon-king function $\psi(x)$.

In Table 1, we provide the analytical definitions of some above mentioned outlier statistics. In practical applications, the block, inward, and outward testing methods can be freely combined with the six statistical measures listed in Table 1 to accommodate different outlier testing scenarios. Overall, the outward testing method may overestimate the number of outliers and perform poorly when there is only one outlier. Both inward and outward testing methods show similar performance when dealing with multiple scattered outliers. The inward testing method using the MRS statistic is more computationally convenient than the outward method, results in fewer false positives, and may even be more effective in identifying single or multiple scattered outliers. In the outward testing method, the MS statistic performs better, while the robustness-modified MRS shows similar performance. In cases of clustered outliers, the mixture testing method is the most convenient and effective (Wheatley and Sornette, 2015).

## 2 Dragon-kings versus characteristic earthquakes

The concept of dragon-kings, when applied to seismicity, significantly overlaps with the notion of characteristic earthquakes. Both ideas involve the occurrence of exceptionally large and relatively predictable seismic events that stand out from the typical distribution of earthquake magnitudes. Dragon-kings describe extreme outliers arising from unique mechanisms, while characteristic earthquakes refer to repetitive, large events specific to certain fault segments. Together, these concepts highlight the importance of understanding extreme seismic phenomena for improved earthquake prediction and hazard assessment. In this section, we identify the similitudes and emphasize the important differences between these two concepts.

In seismology, the concept of "characteristic earthquake" refers to a theoretical model that suggests certain faults or fault segments tend to produce earthquakes of similar size and characteristics on specific faults or fault segments (Schwartz and Coppersmith, 1984; Wesnousky, 1994). In some versions of the concept, the characteristic earthquakes are also thought to exhibit some temporal regularity, occurring at fairly regular intervals, but this has been falsified with the *M* 6.0 Parkfield earthquake of September 28, 2004, which occurred more than 10 years after the upper bound of the supposed 95% confidence interval previously estimated based on the sequence of six previous *M* 6.0 earthquakes since the 1850s. The idea is that these earthquakes have a similar magnitude and rupture area, implying a repeated, predictable seismic behavior on the fault in terms of their rupture type, localization and size. The magnitude consistency of the characteristic earthquakes is hypothesized because the same segment of the fault is supposed to release accumulated stress in a similar manner each time it ruptures over its whole area. Characteristic earthquakes are supposed to occur because there are well-defined fault segments, and each segment can have its own characteristic earthquake, which may be distinct in terms of magnitude and rupture length from those on adjacent segments. In addition to the two founding papers cited above, evidence for characteristic earthquakes is presented by Sachs et al. (2012), which considers a sub-catalog featuring all 1972-2009 events that occurred in a zone fitting with the location of aftershocks of the 2004 Parkfield event in central California. The extrapolation of the Gutenberg-Richter law to large magnitudes "predicts" that the largest event should have a magnitude approximately equal to $M = 5.65$, whereas the largest event that occurred on September 28, 2004 had $M = 5.95$. This discrepancy is taken as the "characteristic" Parkfield event. Amitrano (2012), analyzing the empirical distribution of earthquake energies in California from 1990 to 2010 using the ANSS catalog, finds that the magnitude distribution for events larger than $M = 6$ shows a significant deviation from the power-law. This discrepancy suggests that these larger events could be considered characteristic earthquakes.

This conclusion is however highly controversial and deserves much more scrutiny than what has been offered by the above research. Indeed, the characteristic earthquake hypothesis has faced significant challenges during previous empirical testing. Kagan and Jackson (1991) evaluated McCann et al.'s (1979) model by examining

post-1979 seismic events and found that large earthquakes occurred more frequently in areas deemed to have low seismic potential by McCann et al. (1979). Nishenko (1989, 1991) refined the characteristic earthquake model and applied it to various zones for prospective testing at a high confidence level. However, Kagan and Jackson (1995) observed that earthquake data post-1989 did not support Nishenko's refinements, a finding corroborated by Rong et al. (2003). Jackson and Kagan (2006) provided a detailed critique of the characteristic model and its testing. The hypothesis was vigorously debated in the 1999 Nature discussion on earthquake prediction, with many seismologists either overlooking the test results or attributing failures to specific faults in Nishenko's application. Field and Page (2011) proposed what they call a "relatively objective and reproducible" inverse methodology for determining the rate of different ruptures on a fault or fault system in order to remove the arbitrariness of previous approaches regarding fault segmentation assumptions. For the southern San Andreas fault, their results imply that a Gutenberg-Richter distribution is consistent with the data available for this fault without the need for characteristics earthquakes. But they also point out that "more work is needed to test the robustness of this assertion". Main and Naylor (2012) reviewed the evidence for characteristic earthquakes in comparison with material failure in the laboratory and volcanic eruptions. Main and Naylor (2012) also concluded that the evidence for characteristic earthquakes is inconclusive, though their existence cannot be entirely dismissed. Given that previous applications of the characteristic earthquake hypothesis have consistently resulted in clear failures and no new testable models have been proposed since the mid-1990s, one could thus conclude that this hypothesis is unsupported.

Page et al. (2011) and Main and Naylor (2012) emphasized that power law distributions inherently allow for significant fluctuations, which can sometimes lead to very large values for the largest events within a finite sample. This variability means that the largest event from a power law distribution can sometimes be 10 times larger or more than the second largest event. Such a scenario could misleadingly suggest the presence of a dragon-king, even though it might just be a statistical fluctuation. For instance, in the study by Sachs et al. (2012), the largest observed event was only twice the size of the largest predicted event, highlighting the challenges in distinguishing genuine characteristic events from deviations in the Gutenberg-Richter distribution. The seismicity analyzed by Sachs et al. (2012), which aligns with the assumed seismic cycle duration for the fault segment, could be dominated by aftershocks of the 2004 event. This is where Bäth's law becomes relevant in seismology, quantifying the typical size gap between the largest event and the second largest event from a power law distribution. Using the self-excited Hawkes conditional Poisson model (the epidemic-type aftershock sequence ETAS model in seismology), Helmstetter and Sornette (2003) demonstrated that the magnitude gap of about 1.2 between a main shock and its largest aftershock can be precisely explained by combining the power law distribution of earthquake energies with the scaling law for the dependence of fertility (the number of aftershocks as a function of the magnitude of the main shock). Main and Naylor (2012) caution that great care must be taken when attributing significance to potential "dragon-kings" identified by a large magnitude gap or by comparing observed frequencies with trends extrapolated from distributions at smaller magnitudes. The very search for dragon-kings may introduce a hidden bias, which needs to be considered in statistical tests of the significance of a large magnitude gap.

## 3 Possible physical mechanisms of dragon-king earthquake events

We refer to those large earthquake events whose seismogenic mechanisms are entirely different from those of smaller earthquakes as dragon-king earthquake events. These extreme events are statistically manifested as outliers deviating from the power law e.g., Gutenberg-Richter law. This section presents a few possible physical mechanisms at the origin of dragon-king earthquake events.

### *3.1 Strong fault coupling mechanism*

The strong fault coupling mechanism belongs to the class of partial global synchronization mechanisms previously mentioned in section 1.2 that leads to dragon-king events. The spring-slider model (Burridge and Knopoff, 1967; Carlson and Langer, 1989a; 1989b) is used to illustrate the role of coupling between faults and their sliding behaviors. Numerical simulations by Sachs et al. (2011) have shown that increasing the stiffness of the springs in the spring-slider system produces larger sliding events. Above a certain high stiffness, system-wide slider slip events (i.e., dragon-king earthquake events) occur, and these events no longer follow the power law distribution of the smaller events. The spring-slider system can be viewed as a network of coupled threshold oscillators of relaxation (Schmittbuhl et al., 1993; 1996; Sornette et al., 1994; Gil and Sornette, 1996; Osorio et al., 2010). When the coupling strength of the system (quantified by spring stiffness) is sufficiently large, the oscillators tend to synchronize into system-wide extreme events. Seismological evidence from subduction zones also shows that, when the normal stress is large, faults are highly coupled, and earthquake ruptures are highly correlated with these high-coupling fault areas (Scholz, 2010; Scholz and Campos, 2012). In other words, once there is high coupling across the entire fault system, it is possible for super-large asperities to occur with rupture spreading over multiple fault segments, producing dragon-king earthquake events.

*3.2 Finite fault rupture mechanism*

Some seismological studies suggest that, although smaller earthquakes on faults usually follow a power law distribution (i.e., the Gutenberg-Richter law), most deformation in fault areas is associated with large-magnitude, quasi-periodic characteristic earthquakes (Wesnousky, 1994; Dahmen et al., 1998). Characteristic earthquakes are significant earthquakes that can rupture entire finite fault systems or fault segments and are considered the most representative dragon-king events (Sachs et al., 2012). These earthquakes primarily occur on plate boundary faults (Wesnousky, 1994), corresponding to ruptures that penetrate the entire depth of the brittle crust. Differences in the scale of plate boundary and intra-plate faults are the main reasons characteristic earthquakes differ from other earthquakes. Studies have also found that characteristic earthquakes mainly occur on late Quaternary faults or active faults throughout the earthquake cycle (Ishibe and Shimazaki, 2008; Scholz, 1997). Additionally, fault zones with fast slip rates (> 20 mm/year) also commonly exhibit characteristic earthquakes, and multiple fault ruptures are very common in characteristic earthquakes (Thingbaijam et al., 2024). Characteristic earthquakes statistically manifest as outliers deviating upward from the Gutenberg-Richter law, typifying dragon-king earthquake events.

Moreover, earthquake catalogues that statistically deviate downward from the Gutenberg-Richter law also exist. In these catalogues, the large and small ends of earthquakes follow different slopes (i.e., *b*-values) of the Gutenberg-Richter law (Pisarenko and Sornette, 2004; Pacheco et al., 1992; Scholz, 1997; Triep and Sykes, 1997). Various analyses suggest that the large end actually belongs to another distribution, such as the gamma distribution (also called the Kagan distribution; Kagan, 1991a; see also Chapter 3.3.5 in Sornette (2006) that provides a mathematical derivation of the gamma distribution). The gamma distribution introduces the concept of a corner magnitude, positing that below this corner magnitude, the distribution follows the Gutenberg-Richter law, while above it, it follows an exponential distribution (Kagan, 2002a; 2002b; Serra and Corral, 2017). This truncated distribution is believed to be related to the finite thickness of the brittle crust, meaning that once a rupture penetrates the entire width of a fault, further development of the rupture can only continue in a two-dimensional along-strike direction. This differential rupture propagation mechanism contributes to the statistical differences between large and small earthquakes, but it does not lead to dragon-king earthquake events.

However, some studies and analyses consider that the aforementioned statistical outliers might just be biases of statistical analyses, and thus question the authenticity of characteristic earthquakes (Page et al., 2011; Kagan, 1993; Kagan et al., 2012; Main, 2000). Although the concept of characteristic earthquakes remains controversial academically, it is still widely used in seismic hazard analysis and earthquake engineering practice to assess potential disaster risks (e.g., Zafarani et al., 2019). Therefore, introducing dragon-king theory into seismicity analysis provides a principled framework to consider what are characteristic earthquakes, leading to systematic and effective quantitative analysis tools to rigorously test the concept of characteristic earthquakes.

*3.3 Complex fault rupture structures*

Another potential mechanism for dragon-king earthquakes is multi-segment complex rupturing, manifesting as intermittent cascading rupture behavior. Paleoseismological data confirm that many major earthquakes are produced by cascading ruptures across multiple faults (Scholz, 2010). According to Vere-Jones's (1977) fault branching rupture model, a rupturing fault segment can trigger other fault segments to rupture, thereby generating earthquakes through a cascading process. Let us define the branching ratio as the average number of fault segments to rupture via direct triggering by each rupturing fault segment. When the branching ratio is less than 1, the fault system is in a subcritical state, and the rupture will self-arrest. When the branching ratio is greater than or equal to 1 (critical or supercritical), there is a finite positive probability for the rupture will propagate infinitely through the fault system until it penetrates the brittle crust or encounters a barrier. This situation is related to characteristic earthquakes and corresponds to dragon-king earthquake events of the entire fault system scale.

Moreover, the physics of energy radiation shows that the power radiated by a moving fault (i.e. the energy per unit time radiated outward) is directly proportional to the square of the fault displacement acceleration, i.e., $P \propto \ddot{u}^2$, where $u$ is the slip displacement (Johansen and Sornette, 1999). This means that the presence of asperities increases the efficiency of energy release from faults. Thus, the geometric complexity of faults might lead to different acceleration patterns; in other words, more complex types of fault ruptures might enhance the energy radiation rate, ultimately leading to large-scale energy release forming dragon-king earthquake events.

### 3.4 Unstable run-away rupture mechanism

In recent years, Professor Xiaofei Chen's research group has conducted rupture dynamics simulations based on the slip-weakening friction law, discovering that the rupture stress drop $T_u$, dynamic stress drop $T_e$, critical slip distance $D_c$, effective radius of nucleation asperities $R_a$, and the medium's shear modulus $\mu_0$ together determine the type of earthquake rupture, namely, self-arresting earthquakes, unstable run-away ruptures (i.e., sub-shear and super-shear earthquakes), and self-arresting slow earthquakes (Wen et al., 2018; Wei et al., 2021). Sornette et al. (2024) recently proposed a statistical thermodynamics method to understand the nucleation of different earthquake types, suggesting that self-arresting earthquakes can be likened to stable nucleation phenomena in classical nucleation theory, such as the self-arresting growth of cracks or droplets. Conversely, if the nucleation area exceeds the critical size, the rupture will continue to expand until it encounters a barrier and is forced to stop. This phenomenon is similar to unstable nucleation in classical nucleation theory, revealing the coexistence of stable and unstable ruptures. Unstable run-away ruptures (sub-shear and super-shear earthquakes) are influenced by positive feedback, i.e., as the rupture develops, the released elastic energy further promotes the continuation of the rupture, and friction is unable to stop the continuation, requiring additional dissipation mechanisms through seismic wave energy radiation to achieve dynamic balance. In contrast, in self-arresting earthquakes, friction and seismic energy radiation together act as dissipation mechanisms to terminate the rupture. Clearly, self-arresting earthquakes and unstable run-away ruptures have entirely different mechanisms, with the former being regular earthquakes that follow the Gutenberg-Richter law, while the latter are outlier dragon-king earthquake events. Under Sornette et al.'s (2024) nucleation perspective, unstable run-away ruptures belonging to dragon-king earthquake events should be rarer compared to self-arresting regular earthquakes and of larger sizes in ensemble. In other words, the occurrence probability and size of dragon-king earthquake events are related to the probability of the presence of barriers that prevent their unstable ruptures. It remains to be determined if these two populations can be disentangled from seismic catalogues.

### 4 Prospects for the application of dragon-king theory in seismology

Applying dragon-king theory to seismology can help to systematically quantify and analyze the robustness and domains of validity of many power law distributions currently recognized in seismology, and whether their tails contain outliers (i.e., dragon-king events). By utilizing dragon-king theory, it is possible to revisit the classification of earthquake rupture types from a statistical testing perspective, greatly enriching the analytical tools and research content of statistical seismology. This section will explore the prospects of applying dragon-king theory to seismicity analysis.

## 4.1 Selection of spatiotemporal windows for analysis

Unlike other seismicity analysis methods, dragon-king theory, primarily focused on right tail events (i.e., extreme events), does not require strict completeness of earthquake catalogs. The analysis of seismicity inevitably involves setting the corresponding spatiotemporal window ranges. Therefore, when applying dragon-king theory to seismicity analysis, it is necessary first to select one or a series of suitable spatiotemporal windows to determine the research materials to be analyzed. As illustrated in Figure 2, the choice of spatiotemporal windows largely determines the shape of the frequency-magnitude distribution and, considering that the selection of spatiotemporal windows is significantly influenced by artificial factors, the testing of dragon-king earthquake events needs to exclude these influences to avoid identifying false dragon-king earthquake events. In other fields of application, dragon-king theory does not require such considerations; in applications such as landslides, finance, and sociology, the selection mostly involves choosing time windows.

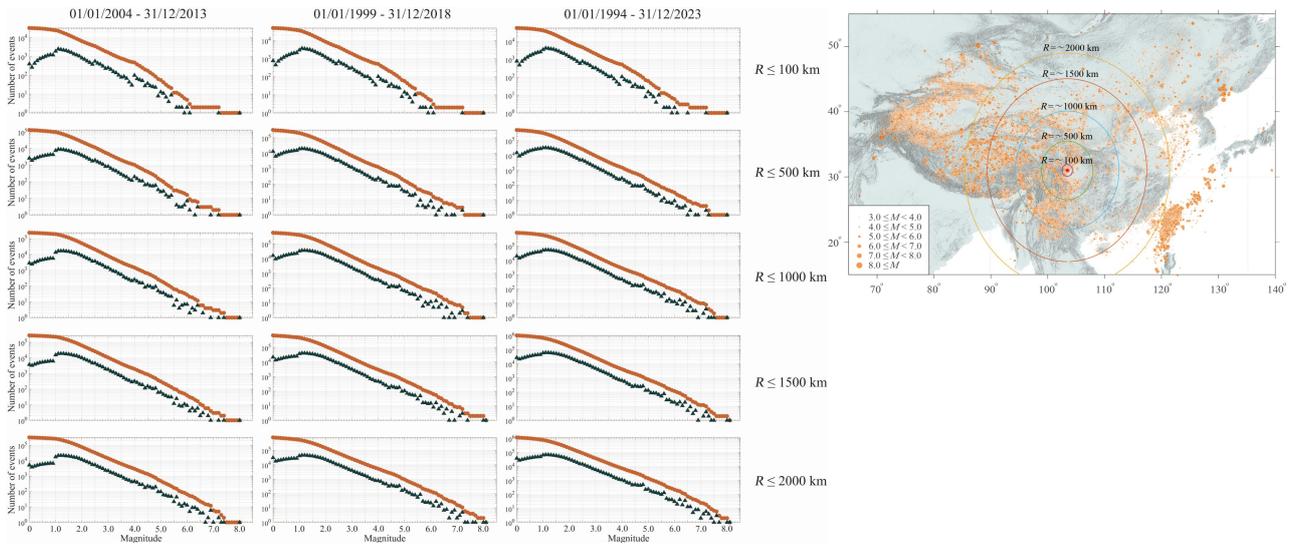

Fig. 2 The relationship between earthquake frequency-magnitude distribution and various spatiotemporal window combinations. Centered on the epicenter and time of the 2008 Sichuan Wenchuan $M$ 8.0 earthquake, earthquakes are selected within epicenter distances of $R \leq$ 100 km, 500 km, 1000 km, 1500 km, and 2000 km (from top to bottom) and within a total of 10 years, 20 years, and 30 years before and after the earthquake occurrence (from left to right). The dark green dots represent the probability density function (PDF), and the orange dots represent the complementary cumulative distribution function (CCDF). A simple observation reveals that, except in the 30-year and 2000 km spatiotemporal window, the Wenchuan earthquake exhibits outlier characteristics of dragon-king earthquake events. Additionally, the 2013 Sichuan Lushan $M$ 7.2 (local magnitude) earthquake also appears as an outlier in smaller spatial windows ($R \leq$ 500 km).

A possible approach is to identify a fault and consider combinations of multiple time and spatial resolutions to examine the robustness of detecting dragon-king earthquake events (as shown in Figure 2). An important bottleneck for the operational definition of an earthquake is how to independently and objectively determine a fault on which the earthquake is supposed to occur, given that faults are elements of approximately self-similar or hierarchical fault networks (Ouillon et al., 1996), and also often consist in the aggregation of smaller-size sub-faults. New clustering techniques that reverse engineer fault networks from seismicity (Ouillon et al., 2008; Ouillon and Sornette, 2011) may provide the basis to treat the earthquake-fault chicken-egg problem consistently (Sornette, 1991). Identifying fault segments is still underlying most of seismic hazard methods. As an illustration, the fault segmentation method was still underlying the fault segmentation used for the hazard calculations in the region at the time of the March 2011 Tohoku earthquake. This approach led to a dramatic under-estimation of the seismic hazard in the Tohoku region, so that an earthquake of $M_W = 9.0$ was indeed thought to be essentially impossible in that region. In fact, the 2011 Tohoku earthquake involved a complex rupture process, including the activation of at least three main fault segments.

Numerical simulation experiments can also be integrated to determine the boundaries of the detectable domain for dragon-king earthquake events. It is necessary to demonstrate that the phenomenon of dragon-king earthquake events remains robust even as the time window is extended. Another potential method is to test for dragon-king earthquake events within the family of earthquake clusters. Currently, many effective statistical seismology tools are available that can cluster earthquake sequences into branching processes to construct the lineage of seismicity triggering (such as the ETAS model; Ogata, 1988; 1998; Ogata and Zhuang, 2006; Hawkes and Oakes, 1974; Kagan, 1991b; Sornette and Werner, 2005; Li et al., 2024).

*4.2 Selection of variables for testing*

The Gutenberg-Richter law describes the frequency distribution of earthquake magnitudes, which can be characterized in different ways, including seismic moment/moment magnitude ($M_W$), radiated energy/energy magnitude ($M_e$), and surface wave magnitude, among others. Although multiple earthquakes may share the same moment magnitude ($M_W$), their energy magnitudes ($M_e$) can differ, reflecting the different dynamic characteristics of the source rupture processes, with slow earthquakes being the most extreme examples (Ide et al., 2007; Obara and Kato, 2016). Moment magnitude measures the change in static elastic energy in the crust before and after an earthquake, whereas energy magnitude describes the total radiated energy at different frequencies resulting from the rupture dynamics, with differences between these measurements stemming from the varying acceleration and deceleration during the rupture process. Research has shown that earthquakes with similar moment magnitudes (and even similar source mechanisms) can have vastly different radiated energies (Choy and Boatwright, 2009; Song et al., 2016). It is worth considering that dragon-king earthquake events might occur in the form of $M_e$ rather than $M_W$ or via the analysis of multivariate objects such as a vector of different magnitudes and other physical characteristics for each earthquake. This requires enriching significantly the methods of outlier detections to the multivariate distribution domain.

*4.3 Beyond the Gutenberg-Richter law*

In fact, the testing framework of dragon-king theory is applicable not only to the Gutenberg-Richter magnitude-frequency relationship but also to other power law distributions in seismology (such as the frequency distribution of fault rupture sizes, fractal/multifractal/hierarchical scaling features of seismicity and fault systems, rank-ordering of earthquakes, frequency distribution of seismicity rates in spatiotemporal grids, frequency distribution of intervals between large earthquakes and epicentral distances, cumulative frequency distribution of earthquake fault slips, and so on). Rigorously testing their power law behavior and investigating the presence of dragon-king events within them opens a rich research agenda that is likely to reveal important new insights.

As seen in Figure 2, outliers in earthquakes can appear as single, multiple, or even clusters of outliers, and the occurrence of these situations largely depends on the choice of spatiotemporal windows. Given that no targeted tests have been conducted in seismology to date, testing strategies and statistics for detecting dragon-king earthquake events, as well as developing a framework that can dynamically adapt and select the optimal testing strategy based on specific data characteristics, are promising directions for future efforts. Moreover, it has been found in practice that the Gutenberg-Richter law is sometimes not a pure power law distribution but a truncated power law distribution with an exponential tail (Kagan, 1994; Main, 1996; Sornette and Sornette, 1999) or can even present a two-branched structure when considering the cluster of events directly triggered by a given earthquake (Vere-Jones, 2005; Saichev and Sornette, 2005; Nandan et al., 2019; 2022). Therefore, this may also need to be considered when constructing the null hypothesis for tests.

**5 Outlook**

In future seismological research and earthquake risk reduction practices, understanding and predicting large earthquakes continue to pose significant challenges. Although the mechanisms behind large earthquakes may be similar to those of smaller earthquakes in some respects, the differences in distributions as well as in a number of physical properties between large and small earthquakes in specific regions suggest that large earthquakes may have unique characteristics in their genesis and rupture mechanisms. We proposed to formulate the question of whether large to great earthquakes are physically and/or statistically different from smaller earthquakes within the framework of the dragon-king theory. Based on this conceptual structure, we introduced systematic and rigorous framework for statistical testing of endogenous outliers, and outlined the possible physical mechanisms that produce such endogenous outlier behaviors. Integrating with seismological theories, we discussed several fault rupture mechanisms that could generate endogenous earthquake outliers (i.e., dragon-king earthquake events) and looked forward to the operability of applying dragon-king theory in seismological practice.

We noted that incorrect or incomplete statistical analysis may produce artificially false dragon-king results. Therefore, every suspected dragon-king earthquake event result should be continuously scrutinized with skepticism, and analyses should be refined and expanded. Dragon-king theory posits that the mechanisms producing dragon-king events have some unique characteristics and may produce precursors different from those of other events in the same system prior to their occurrence, thereby endowing dragon-king events with better predictability than their smaller siblings. Dragon-king theory emphasizes the importance of dynamically monitoring these precursor phenomena and of developing real-time modeling and analysis of risks. This also aligns with our plans to develop earthquake forecasting systems in China and globally that integrate multi-type precursor observations and interdisciplinary collaborative research (Mignan et al., 2021; Li et al., 2023).

Acknowledgements

The authors would like to thank Yang Zang from the China Earthquake Networks Center for his assistance in accessing the Chinese earthquake catalog data. We are also grateful to Professor Jianchang Zhuang for his valuable discussions. Additionally, we appreciate the support from Xinyi Wang, Qingyuan (Tsingyuan) Zhang, Zhongpu (Oscar) Qiu, and Chunmei Yang in conducting the literature search and organizing the materials. This research was supported by the Guangdong Basic and Applied Basic Research Foundation (2024A1515011568), the National Natural Science Foundation of China (U2039202 and U2039207), the Shenzhen Science and Technology Innovation Commission (GJHZ20210705141805017), and the Center for Computational Science and Engineering at Southern University of Science and Technology.

# 地震统计物理研究的新视角：
# "龙王"理论及"龙王"地震事件

李佳威 [1]，宋笛牒 [1*]，吴忠良 [1,2]，李杭威 [1]

1 南方科技大学前沿与交叉科学研究院风险分析预测与管控研究院，深圳　518055
2 中国地震局地震预测研究所，北京　100036

**摘要**：系统地定量考察大地震发震机理是否特殊，也许不仅能够深化对断层破裂和地震活动规律的理解，而且有望促进大地震可预测性研究和提升防灾减灾策略的有效性。本文作者之一于 2009 年提出的"龙王"理论，旨在为那些内源性极端离群值（即"龙王"事件）提供一个量化的识别和检验框架。该理论为解释、预测和管控这些罕见但影响巨大的事件提供了非常有价值的分析工具。本文讨论了将其应用于地震学的可行性，提出"龙王"地震事件可作为古登堡-里克特定律的离群值来识别，并探讨了可能导致这些特殊事件产生的若干地震学机制。尽管"龙王"理论在地震学的具体应用还面临诸多实际挑战，但其有望显著丰富统计地震学的研究内容。通过从统计检验的视角重新系统性审视地震破裂类型的分类，并将这些认识与其背后的物理机制相结合，本文呈现的内容将可以极大提升地震统计物理研究领域的分析手段和研究深度。

**关键词**：地震统计物理；"龙王"事件；"龙王"理论；地震发震机制；统计地震学；离群值检验

**基金项目**：广东省基础与应用基础研究基金（2024A1515011568）；国家自然科学基金（U2039202、U2039207）；深圳市科技创新委员会（GJHZ20210705141805017）；南方科技大学计算科学与工程中心

Supported by the Guangdong Basic and Applied Basic Research Foundation (Grant No. 2024A1515011568), the National Natural Science Foundation of China (Grant no. U2039202 and U2039207), Shenzhen Science and Technology Innovation Commission (Grant no. GJHZ20210705141805017), and the Center for Computational Science and Engineering at Southern University of Science and Technology.

**第一作者**：李佳威（1992-），男，博士，主要从事地震学研究. E-mail: lijw3@sustech.edu.cn
***通信作者**：宋笛牒 Didier Sornette (1957-)，男，讲席教授/院士，主要从事地震学、统计物理、复杂系统和风险分析等研究. E-mail: dsornette@ethz.ch